\documentstyle[12pt]{article}
\newcommand{\nc}{\newcommand}
\nc{\renc}{\renewcommand}

\nc{\etal}{\mbox{\it et al. }}
\nc{\ie}{{\it i.e.}}
\nc{\eg}{{\it e.g.}}

\renc{\thefootnote}{\arabic{footnote}}
\nc{\capt}[1]{{\bf Figure.} {\small\sl #1}}

\nc{\eqs}[2]{\mbox{Eqs.~(\ref{#1},\,\ref{#2})}}
\nc{\eq}[1]{\mbox{Eq.~(\ref{#1})}}

\nc{\figs}[2]{\mbox{Figs.~(\ref{#1},\,\ref{#2})}}
\nc{\fig}[1]{\mbox{Fig~.(\ref{#1})}}

\nc{\tag}[1]{\label{#1} \marginpar{{\footnotesize #1}}}
\nc{\mtag}[1]{\label{#1} \mbox{\marginpar{{\footnotesize #1}}}}
\renc{\baselinestretch}{1.2}
\newlength{\overeqskip}
\newlength{\undereqskip}
\nc{\be}[1]{\begin{equation} \mbox{$\label{#1}$}}
\nc{\bea}[1]{\begin{eqnarray} \mbox{$\label{#1}$}}
\nc{\Section}[2]{\section{#2}\label{#1}}
\nc{\Bibitem}[1]{\bibitem{#1}}
\nc{\Label}[1]{\label{#1}}

\nc{\eea}{\vspace{\undereqskip}\end{eqnarray}}
\nc{\ee}{\vspace{\undereqskip}\end{equation}}
\nc{\bdm}{\begin{displaymath}}
\nc{\edm}{\end{displaymath}}
\nc{\dpsty}{\displaystyle}
\nc{\bc}{\begin{center}}
\nc{\ec}{\end{center}}
\nc{\ba}{\begin{array}}
\nc{\ea}{\end{array}}
\nc{\bab}{\begin{abstract}}
\nc{\eab}{\end{abstract}}
\nc{\btab}{\begin{tabular}}
\nc{\etab}{\end{tabular}}
\nc{\bit}{\begin{itemize}}
\nc{\eit}{\end{itemize}}
\nc{\ben}{\begin{enumerate}}
\nc{\een}{\end{enumerate}}
\nc{\bfig}{\begin{figure}}
\nc{\efig}{\end{figure}}
\nc{\arreq}{&\!=\!&}
\nc{\arrmi}{&\!-\!&}
\nc{\arrpl}{&\!+\!&}
\nc{\arrap}{&\!\!\!\approx\!\!\!&}
\nc{\non}{\nonumber\\*}
\nc{\align}{\!\!\!\!\!\!\!\!&&}

\def\lsim{\; \raise0.3ex\hbox{$<$\kern-0.75em
      \raise-1.1ex\hbox{$\sim$}}\; }
\def\gsim{\; \raise0.3ex\hbox{$>$\kern-0.75em
      \raise-1.1ex\hbox{$\sim$}}\; }
\nc{\DOT}{\hspace{-0.08in}{\bf .}\hspace{0.1in}}
\nc{\Laada}{\hbox {$\sqcap$ \kern -1em $\sqcup$}}
\nc\loota{{\scriptstyle\sqcap\kern-0.55em\hbox{$\scriptstyle\sqcup$}}}
\nc\Loota{{\sqcap\kern-0.65em\hbox{$\sqcup$}}}
\nc\laada{\Loota}
\nc{\qed}{\hskip 3em \hbox{\BOX} \vskip 2ex}

\nc{\real}{{\rm I \! R}}
\nc{\Z}{{\sf Z \!\!\! Z}}
\nc{\complex}{{\rm C\!\!\! {\sf I}\,\,}}
\def\bigid{\leavevmode\hbox{\small1\kern-3.8pt\normalsize1}}
\def\id{\leavevmode\hbox{\small1\kern-3.3pt\normalsize1}}
\nc{\slask}{\!\!\!/}
\nc{\bis}{{\prime\prime}}
\nc{\pa}{\partial}
\nc{\na}{\nabla}
\nc{\ra}{\rangle}
\nc{\la}{\langle}
\nc{\goto}{\rightarrow}
\nc{\swap}{\leftrightarrow}

\nc{\EE}[1]{ \mbox{$\cdot10^{#1}$} }
\nc{\abs}[1]{\left|#1\right|}
\nc{\at}[2]{\left.#1\right|_{#2}}
\nc{\norm}[1]{\|#1\|}
\nc{\abscut}[2]{\Abs{#1}_{\scriptscriptstyle#2}}
\nc{\vek}[1]{\mbox{\bf #1}}
\nc{\veks}[1]{\mbox{\scriptsize\bf #1}}
\nc{\vekk}[1]{\mbox{\boldmath #1}}
\nc{\integral}[2]{\int\limits_{#1}^{#2}}
\nc{\inv}[1]{\frac{1}{#1}}
\nc{\dd}[2]{{{\partial #1}\over{\partial #2}}}
\nc{\ddd}[2]{{{{\partial}^2 #1}\over{\partial {#2}^2}}}
\nc{\dddd}[3]{{{{\partial}^2 #1}\over
	{\partial #2 \partial #3}}}
\nc{\dder}[2]{{{d #1}\over{d #2}}}
\nc{\ddder}[2]{{{d^2 #1}\over{d {#2}^2}}}
\nc{\dddder}[3]{{d^2 #1}\over
	{d #2 d #3}}
\nc{\dx}[1]{d\,^{#1}x}
\nc{\dy}[1]{d\,^{#1}y}
\nc{\dz}[1]{d\,^{#1}z}
\nc{\dl}[1]{\frac{d\,^{#1}l}{(2\pi)^{#1}}}
\nc{\dk}[1]{\frac{d\,^{#1}k}{(2\pi)^{#1}}}
\nc{\dq}[1]{\frac{d\,^{#1}q}{(2\pi)^{#1}}}

\nc{\cc}{\mbox{$c.c.$ }}
\nc{\hc}{\mbox{$h.c.$ }}
\nc{\cf}{cf.\ }
\nc{\erfc}{{\rm erfc}}
\nc{\Tr}{{\rm Tr\,}}
\nc{\tr}{{\rm tr\,}}
\nc{\pol}{{\rm pol}}
\nc{\sign}{{\rm sign}}
\nc{\bfT}{{\bf T }}

\nc{\cA}{{\cal A}}
\nc{\cB}{{\cal B}}
\nc{\cD}{{\cal D}}
\nc{\cE}{{\cal E}}
\nc{\cG}{{\cal G}}
\nc{\cH}{{\cal H}}
\nc{\cL}{{\cal L}}
\nc{\cO}{{\cal O}}
\nc{\cT}{{\cal T}}
\nc{\cN}{{\cal N}}
\nc{\rvac}[1]{|{\cal O}#1\rangle}
\nc{\lvac}[1]{\langle{\cal O}#1|}
\nc{\rvacb}[1]{|{\cal O}_\beta #1\rangle}
\nc{\lvacb}[1]{\langle{\cal O}_\beta #1 |}
\nc{\bb}{\bar{\beta}}
\nc{\bt}{\tilde{\beta}}
\nc{\ctH}{\tilde{\cal H}}
\nc{\chH}{\hat{\cal H}}
\nc{\lagr}{{\cal L}}
\nc{\dsub}[1]{\partial_{#1}}
\nc{\dsup}[1]{\partial^{#1}}
\nc{\av}[1]{\langle #1 \rangle}
\nc{\ordo}[1]{{\cal O}(#1)}
\nc{\conj}[1]{\overline{#1}}
\nc{\orp}{\omega_r^+}
\nc{\orm}{\omega_r^-}
\nc{\1}{\aa}
\nc{\2}{\"{a}}
\nc{\3}{\"{o}}
\nc{\4}{\AA}
\nc{\5}{\"{A}}
\nc{\6}{\"{O}}
\nc{\al}{\alpha}
\nc{\g}{\gamma}
\nc{\Del}{\Delta}
\nc{\e}{\epsilon}
\nc{\eps}{\epsilon}
\nc{\lam}{\lambda}
\nc{\om}{\omega}
\nc{\Om}{\Omega}
\nc{\ve}{\varepsilon}
\nc{\mn}{{\mu\nu}}
\nc{\k}{\kappa}
\nc{\vp}{\varphi}

\nc{\advp}[3]{{\it  Adv.\ in\ Phys.\ }{{\bf #1} {(#2)} {#3}}}
\nc{\annp}[3]{{\it  Ann.\ Phys.\ (N.Y.)\ }{{\bf #1} {(#2)} {#3}}}
\nc{\apl}[3]{{\it  Appl. Phys. Lett. }{{\bf #1} {(#2)} {#3}}}
\nc{\apj}[3]{{\it  Ap.\ J.\ }{{\bf #1} {(#2)} {#3}}}
\nc{\apjl}[3]{{\it  Ap.\ J.\ Lett.\ }{{\bf #1} {(#2)} {#3}}}
\nc{\app}[3]{{\it Astropart.\ Phys.\ }{{\bf #1} {(#2)} {#3}}}
\nc{\cmp}[3]{{\it  Comm.\ Math.\ Phys.\ }{{ \bf #1} {(#2)} {#3}}}
\nc{\cqg}[3]{{\it  Class.\ Quant.\ Grav.\ }{{\bf #1} {(#2)} {#3}}}
\nc{\epl}[3]{{\it  Europhys.\ Lett.\ }{{\bf #1} {(#2)} {#3}}}
\nc{\ijmp}[3]{{\it Int.\ J.\ Mod.\ Phys.\ }{{\bf #1} {(#2)} {#3}}}
\nc{\ijtp}[3]{{\it Int.\ J.\ Theor.\ Phys.\ }{{\bf #1} {(#2)} {#3}}}
\nc{\jmp}[3]{{\it  J.\ Math.\ Phys.\ }{{ \bf #1} {(#2)} {#3}}}
\nc{\jpa}[3]{{\it  J.\ Phys.\ A\ }{{\bf #1} {(#2)} {#3}}}
\nc{\jpc}[3]{{\it  J.\ Phys.\ C\ }{{\bf #1} {(#2)} {#3}}}
\nc{\jap}[3]{{\it J.\ Appl.\ Phys.\ }{{\bf #1} {(#2)} {#3}}}
\nc{\jpsj}[3]{{\it J.\ Phys.\ Soc.\ Japan\ }{{\bf #1} {(#2)} {#3}}}
\nc{\lmp}[3]{{\it Lett.\ Math.\ Phys.\ }{{\bf #1} {(#2)} {#3}}}
\nc{\mpl}[3]{{\it  Mod.\ Phys.\ Lett.\ }{{\bf #1} {(#2)} {#3}}}
\nc{\ncim}[3]{{\it  Nuov.\ Cim.\ }{{\bf #1} {(#2)} {#3}}}
\nc{\np}[3]{{\it  Nucl.\ Phys.\ }{{\bf #1} {(#2)} {#3}}}
\nc{\pr}[3]{{\it Phys.\ Rev.\ }{{\bf #1} {(#2)} {#3}}}
\nc{\pra}[3]{{\it  Phys.\ Rev.\ A\ }{{\bf #1} {(#2)} {#3}}}
\nc{\prb}[3]{{\it  Phys.\ Rev.\ B\ }{{{\bf #1} {(#2)} {#3}}}}
\nc{\prc}[3]{{\it  Phys.\ Rev.\ C\ }{{\bf #1} {(#2)} {#3}}}
\nc{\prd}[3]{{\it  Phys.\ Rev.\ D\ }{{\bf #1} {(#2)} {#3}}}
\nc{\prl}[3]{{\it Phys\ Rev.\ Lett.\ }{{\bf #1} {(#2)} {#3}}}
\nc{\pl}[3]{{\it  Phys.\ Lett.\ }{{\bf #1} {(#2)} {#3}}}
\nc{\prep}[3]{{\it Phys\. Rep.\ }{{\bf #1} {(#2)} {#3}}}
\nc{\prsl}[3]{{\it Proc.\ R.\ Soc.\ London\ }{{\bf #1} {(#2)} {#3}}}
\nc{\ptp}[3]{{\it  Prog.\ Theor.\ Phys.\ }{{\bf #1} {(#2)} {#3}}}
\nc{\ptps}[3]{{\it  Prog\ Theor.\ Phys.\ suppl.\ }{{\bf #1} {(#2)} {#3}}}
\nc{\physa}[3]{{\it  Physica\ A\ }{{\bf #1} {(#2)} {#3}}}
\nc{\physb}[3]{{\it  Physica\ B\ }{{\bf #1} {(#2)} {#3}}}
\nc{\phys}[3]{{\it Physica\ }{{\bf #1} {(#2)} {#3}}}
\nc{\rmp}[3]{{\it  Rev.\ Mod.\ Phys.\ }{{\bf #1} {(#2)} {#3}}}
\nc{\rpp}[3]{{\it Rep.\ Prog.\ Phys.\ }{{\bf #1} {(#2)} {#3}}}
\nc{\sjnp}[3]{{\it Sov.\ J.\ Nucl.\ Phys.\ }{{\bf #1} {(#2)} {#3}}}
\nc{\spjetp}[3]{{\it Sov.\ Phys.\ JETP\ }{{\bf #1} {(#2)} {#3}}}
\nc{\yf}[3]{{\it Yad.\ Fiz.\ }{{\bf #1} {(#2)} {#3}}}
\nc{\zetp}[3]{{\it Zh.\ Eksp.\ Teor.\ Fiz.\  }{{\bf #1}  {(#2)} {#3}}}
\nc{\zp}[3]{{\it Z.\ Phys.\ }{{\bf #1} {(#2)} {#3}}}
\nc{\ibid}[3]{{\sl ibid.\ }{{\bf #1} {#2} {#3}}}
\nc{\rf}[1]{(\ref{#1})}
\nc{\nn}{\nonumber \\*}
\nc{\SM}{Standard~Model~}
\nc{\MP}{M_{\rm Pl}}
\nc{\tp}{t_{\rm Pl}}
\begin{document}

{\title{\vskip-2truecm{\hfill {{\small  TURKU-FL-R95-24\\
         \hfill HU-TFT-95-58\\
         \hfill hep-ph/9509359\\
        }}\vskip 1truecm} {\bf Damping rate of neutrinos in the singlet
Majoron model}}


{\author{ {\sc Timo Holopainen $^{1}$ }\\  {\sl\small Department of
Physics, University of Turku,  FIN-20500 Turku, Finland} \\ {\sc Jukka
Maalampi$^{2}$ }\\  {\sl\small Department of Physics, Theory Division,
P.O. Box 9,
FIN-00014 University of  Helsinki, Finland} \\ {\sc Jukka Sirkka$^{3}$ }\\
{\sc \small and}\\ {\sc Iiro Vilja$^{4}$ }\\  {\sl\small Department of
Physics, University of Turku, FIN-20500 Turku, Finland}}}

\maketitle
\vspace{2cm}
\begin{abstract}
\noindent  The damping rate and free path of neutrinos in the singlet
Majoron model have been calculated including both finite  temperature and
symmetry breaking effects. The behaviour of right- and left-handed fermions
are found inherently different. While the damping rates of the left-handed
leptons are essentially model independent, e.g. directly applicable to the
Standard Model, for the right-handed particles
the rates are crucially sensitive to parameters of the scalar sector. In
general, the damping rates are fairly large. The possibility of the
right-handed neutrinos to penetrate deep into the broken phase in the
electroweak phase transition still remains, however, for some parts of the
parameter space.
\end{abstract}
\vfil
\footnoterule {\small $^1$tholopai@utu.fi}; {\small
$^2$maalampi@phcu.helsinki.fi}; {\small  $^3$sirkka@utu.fi}; {\small
$^4$vilja@utu.fi}
\thispagestyle{empty}
\newpage
\setcounter{page}{1}
\Section{Intro}{Introduction}

A particle traversing a thermal medium undergoes continual incoherent
scatterings with plasma particles. Because of these interactions  energy
and momenta of the particle  are not sharply defined but  have a finite
"width",
$\gamma$. The quantity $\gamma$, called the  damping rate, describes the
"decay" of the state (quasiparticle) into the plasma. Only when $\gamma$
is small compared with the energy will the quasiparticle propagate over
meaningful distances in the plasma as a true physical excitation and with
significant physical effects.

The calculation of the damping rate of quasiparticle at zero spatial
momentum, $\gamma_{QCD}(0)$, in QCD plasma has been a widely disputed
issue in literature \cite{contres,Pisarski}. The controversies in the
results obtained for this quantity were solved by Pisarski \cite{Pisarski}
and Braaten and Pisarski
\cite{BP1,BP2} who developed a general method for resumming the higher
loop diagrams which contribute to the damping rate in leading order.

The method of Braaten and Pisarski can be applied not only to QCD but also
to other interactions. In this article we shall use it to  calculate the
damping rates  of low momenta neutrinos
 in the electroweak plasma of the early universe.

We are interested in the determination of the damping rates in connection
to the so-called charge transport mechanism for generating the baryon
asymmetry of the universe during the electroweak phase transition
\cite{ckn}.  This scenario is  based on the assumption that the
electroweak phase transition is of first order.  Bubbles of the broken
phase are created by fluctuations, and they grow and fill the pre-existing
unbroken  phase.
 As a bubble expands a flux of leptons from the unbroken phase hits the
wall separating the phases, and some portion of the flux is reflected. As
a result of violation of lepton number and C- and CP-parities, a net
number of leptons is created in the unbroken phase. This is turned to a
net baryon number by the sphaleron transitions. The produced baryons pass
to the broken region, where baryon number is conserved, resulting in
non-equal numbers of baryons and antibaryons in the universe we live.

 The reflection is generally most efficient at small spatial momenta (that
is why the damping rate at zero momentum is a relevant quantity to study)
for the following reason. When a quasiparticle penetrates to the broken
phase region it has an energy determined by the symmetric phase dispersion
relation, until it begins to see the thermal background of the broken
phase. If a particle has energy which is smaller than the thermal mass it
acquires in the broken phase, it can propagate there only the time
determined by the thermalization rate, after which it is reflected
efficiently. Thermal masses are generally of the order $gT$, i.e.
particles moving with small spatial momenta are reflected efficiently.

If its coherence length $\sim 1/2\gamma$  is small compared with its
effective reflection length, a typical quasiparticle  decays into the
plasma before it is reflected. This would suppress the generation of  the
baryon number making the scenario less viable. Actually, it has been
claimed \cite{gavelaetal,HuetSather} that in the picture where the baryon
number is created, instead of neutrinos, by a reflection of quarks   from
the phase boundary, this is what happens: decoherence effects crucially
diminish the baryon number creation. It is interesting to investigate what
is the situation with neutrinos.  In the present paper we will calculate
the damping rates for left-handed and right-handed neutrinos in the
Majoron model. The consequences of the result with respect to the baryon
number generation will be analyzed in a separate publication
\cite{homasivi}.

The organization of the paper is the following. In Section 2 we will
introduce the model we are working with, the singlet Majoron model. In
Section 3 we derive the effective vertices and propagators, and in Section
4 we apply them to the calculation of the zero momenta damping rates. The
numerical results are presented  in Section 5, and the conclusions drawn in
Section 6.

\Section{model}{The singlet Majoron model}

For evaluating  the neutrino damping rate one has to specify the model. We
will use the singlet Majoron model
\cite{SeMajoronPeruspaperi}, which is an extension of
\SM containing an extra gauge singlet scalar field $S$, as well as gauge
singlet right-handed neutrinos.    The lagrangian of the model is the most
general one with the requirement that it has
a global $B-L$ symmetry. Spontaneous  breaking of this
global symmetry by a non-vanishing vacuum expectation value (vev) of the
singlet scalar, $\langle S\rangle=\bar f$, generates Majorana masses  to
the neutrinos. The see-saw mechanism can be implemented  yielding three
neutrinos which are  very light  compared with the other fermions of the
model and three  neutrinos which are very heavy. It is  usually assumed,
to avoid an ad hoc hierarchy among the energy scales, that
$\bar f$ is  comparable in size with the vev $\langle\Phi\rangle =f$ of the
ordinary doublet  Higgs $\Phi$ responsible on the electroweak symmetry
breaking. The model  can fulfill the Sakharov conditions \cite{Sakke}
for the creation of a cosmological asymmetry  between baryons and
antibaryons, as was demonstrated in \cite{ckn}.

The singlet Majoron model contains the following new  Yukawa interactions
not present in the Standard  Model:
\be{ly}
\lagr_Y=\conj{l}_L\lam_D\tilde\Phi N_R+\conj{N}_R\lam_M S N_L^c.
\ee
Here $\tilde\Phi$ is the conjugate of the Higgs
doublet, $l_L$ is a vector of left-handed  lepton doublets and $N_R$ is  a
vector of right-handed singlet neutrinos.
 The Yukawa coupling matrices $\lam_D$ and $\lam_M$ are in general
non-diagonal  in flavor space.

The first term of \rf{ly} gives rise to Dirac masses and the second
term to Majorana masses for neutrinos. In order to satisfy the
experimental mass limits of the known neutrinos one has to require, given
our  assumption that there is no strong hierarchy between the vevs $f$ and
$\bar f$, that the elements of
$\lam_D$ are much smaller than  the elements of $\lam_M$, so as to make
the  predominantly left-handed neutrinos very light.  A detailed
discussion of the masses, mixings and  phase transitions in the singlet
Majoron  model can be found in \cite{EnqvistKV}. We just note that due to
a large number of free parameters  appearing in the model and a lack of
constraining experimental data, the vacuum expectation value $\bar f$,
singlet scalar mass and Majorana masses can be considered as practically
free  parameters.

\Section{v and p}{ Effective vertices and propagators}
The damping rate is defined as the imaginary part of the solution
$\omega=\omega(k)$ of the dispersion relation
\be{disp1}
\det\left(\Delta_f^{-1}(\omega,\vek k)\right)=\det\left(\omega\gamma^0-
\vek{k}\cdot\mbox{\boldmath $\gamma$}-\Sigma(\omega,\vek k)-m_f\right)=0,
\ee
where $\Delta_f$ is the fermion propagator, $\Sigma$ is the fermion
self-energy and
$m_f$ is the vacuum mass of the fermion. The contributions to the finite
temperature self-energies that one has to consider in order to determine the
zero momentum damping rate of the Majorana neutrinos in the singlet
Majoron model are displayed in Fig. 1.  In this section we shall outline
the computation of the effective propagators and vertices (presented by
the 'blobs') out of which these amplitudes are constructed. In the next
section we will apply the results to evaluate the damping
rates.  In our calculation we will make the following approximations. We
will compute all quantities in leading order in coupling constants,
neglect the Yukawa couplings except the Majorana couplings of the
right-handed neutrinos, and when necessary take subleading terms into
account to avoid problems with infrared divergent graphs.

We will apply the method of  Braaten and Pisarski \cite{BP1} to identify
those  higher-loop diagrams, called hard thermal loops (HTL), which
contribute in leading order and which are necessary for the gauge
invariance  of the damping rate.  By definition, HTL's are those diagrams
which contribution to effective vertices is of the same order in coupling
constants as the contribution of the tree level diagrams. According to
\cite{BP1}, an amplitude
contributing the effective vertices and propagators can be an HTL only
when  its external legs are all  soft, i.e. the momenta of the external
particles are of the order $\sim gT$, where $T$ is temperature and
$g$ is a generic coupling constant.

Let $P=(p_0,\vek p)$ denote the external soft momentum  and
$K=(k_0,\vek k)$ the loop momentum. Whether a given amplitude is an HTL
can be determined by  applying the power counting rules introduced in Ref.
\cite{BP1}: the integration element $\int d^3k$ contributes
$T^3$; the first propagator in the loop and the summation over $k_0$ together
contribute $1/T$, and all other propagators give a contribution
$1/(PT)$ each;
$K^\mu$ in the numerator contributes $T$; $P^\mu$ in the numerator
contributes
$gT$; for loop where internal lines are all bosonic or fermionic there is
an extra factor of $P/T$.

The hard part of an amplitude corresponds to making a linear approximation
in the external momentum $P\sim gT$ in propagators, i.e. neglecting $P^2$
as small compared with $P\cdot K$. The particle masses, which appear  in
the propagators  quadratically,  can be neglected, because they are
generally of the order
$gv$, where $v$ is a generic vacuum expectation value of a symmetry
breaking Higgs boson and $g$ is a generic coupling. An exception  seems to
be scalar masses, which are  lower order in coupling constant,
$m_S^2\sim g v^2$. However, perturbatively in finite temperature one has
$v\sim  g^{1/2} T$, and since the soft scale of scalar lines should be
identified with $\sim g^{1/2}T$, we can  neglect in leading order also the
scalar masses in
loops. For example, calculating the scalar tadpole diagram, which arises
from scalar self-coupling, gives for the scalar a thermal mass $\sim
g^{1/2}T$, which is small compared with the relevant loop momentum.

Let us now move to consider in one loop level the effective two-point,
three-point and four-point functions entering the diagrams of Fig. 1.
There is obviously a great number of loop diagrams contributing, but it
turns out that  just a handful of them are HTL's and thus relevant in
leading order evaluation of the damping rates. We will not  go through
our analysis in full detail diagram by diagram here, but rather just
describe the general methods we have used to discard the amplitudes
yielding a non-leading contribution and to pick up the HTL's.

Let us first consider diagrams for the left-handed neutrinos.
In the case of the gauge boson - neutrino pair three-point function
there are five different types of diagrams, which may be conveniently
classified according to the types of their internal lines. Let F denote
a fermion line, S a scalar line and A a gauge boson line. The diagrams with
configurations FFA and FAA have good infrared behaviour. They are HTL's,
and the calculation of them is presented in the Appendix. The diagrams
FFS and FSS have also good infrared behaviour and they are HTL's, but
because they contain small Yukawa couplings we neglect them. The configuration
FAS has a vertex V(AAS) which originates in the symmetry breaking and which
has with a coupling
proportional to  $g^2 f$. By the power counting rules its contribution
would be $\sim g^2 f/T$, i.e. of higher order than the tree level
contribution. However, due to the behaviour of this graph in the infrared
region, its contribution is larger than given by the pure power counting
argument. To see this, and to introduce our scheme to approximate the
contributions from the infrared integration regions, let us consider
the amplitude FAS in more detail.

Using the standard
imaginary time formalism \cite{kapusta} to perform the summation over
discrete loop  energy (see also Appendix), the diagram FAS may be cast
into the form
\be{loop}
\Gamma({\rm FAS})\sim g^4 f\int\frac{d^3k}{(2\pi)^3} \left[K\cdot\gamma|_
{k_0=\omega_k}
\frac{f_B(\omega_k)}{2\omega_k}\frac 1{(\omega_k+p_1^0)^2-\omega_{k+p_1}^2}
\frac 1{(\omega_k+p_2^0)^2-\omega_{k+p_2}^2}+\dots\right].
\ee
Here $\omega_{k + p}^2=(\vek k + \vek p )^2+m^2$,  $f_B$
is the Bose-Einstein distribution function, the $P_i$'s are some combinations
of the external momenta, and $m$ is the mass of the
loop particle. The external energies $p_i^0$ are here continued  to real
values. If one  makes a linear approximation in the external four-momenta
and neglects the masses in  denominators, the integral in \rf{loop}
becomes infrared divergent and it behaves as
\be{loop2}
\Gamma({\rm FAS})_{\rm IR}\sim g^4 f\int \frac{dk}k \frac T{P^2}
=g^2\frac fT \int\frac{dk}k,
\ee
where $P$ denotes generically the soft external momentum scale.
As the integral is only logarithmically divergent the contribution from the
infrared region is of the order of $g^2\log(g) f/T$. Hence
$\Gamma(FAS)$ is not an HTL in spite of its divergent infrared behaviour.

By generalizing previous example one obtains power counting rules for the
infrared region: the summation over $k_0$ and one propagator contributes
together treating $1/(gT)$; the integration element $d^3k$ contributes
$(gT)^3$;
the loop momentum $K$ is taken to be of the order of the external soft
momentum $P\sim gT$ both in the nominator and in the denominator; and the
Bose-Einstein distribution function contributes $1/g$.

There are five types of diagrams contributing to the scalar-fermion pair
three-point function: FFA, FFS, FAS, FAA and FSS. All but one, FAA, of
these configurations
contain small Yukawa couplings, and hence we shall neglect them.
The diagram FAA is due to the symmetry breaking, and it behaves as
$\sim g^2 f/T$
both in the ultraviolet and infrared regions. Because also the tree level
coupling of the left-handed neutrinos and a scalar is proportional to
small Yukawa coupling, the effective scalar -- neutrino pair vertex can be
omitted for the left-handed neutrinos altogether.

A note concerning the configuration FSS is in order. It contains
a scalar three point vertex arising through the symmetry breaking from the
scalar four-point coupling, generically denoted by $\lam$, of the symmetric
theory. Power counting results in a contribution $\sim g v/T$, where
$v$ is a generic vacuum expectation value parameter. Thus if $v$ is of the
same order than temperature $T$, one should consider the FSS diagram as a HTL.
But here one should be careful to check whether the contribution comes from
the scalar
or from the fermion line (see Eq. \rf{htl2}). By identifying the soft scale
of the scalars with $\lam^{1/2} T$, these contributions are $\sim g_Y^2 v/T$
and $\sim \lam^{1/2}g_Y v/T$, respectively. Here $g_Y$ is a generic Yukawa
coupling. One can check that also the
ultraviolet region gives a subleading contribution and hence FSS diagram
is not an HTL. Formally the same result is achieved if $\lam^{1/2}$ is
considered as the perturbation parameter of the scalar sector.

With the above note in mind one can check that from the 13 possible diagrams
of the two gauge boson -- neutrino pair four-point function only three,
FAAA, FFAA and FFFA, contribute. The calculation of these is outlined in the
Appendix. There are five diagrams which have vertices arising from
the symmetry breaking: FFAS; two diagrams of the type FASS;
two diagrams of the type FAAS
(these are neglected because they are not HTL's); two non-HTL diagrams
with vertices of the symmetric theory: FAA and FSS; and three diagrams
containing small Yukawa couplings: FSSS, FFSS and FFFS.

Further one can check that none of the 13  possible diagrams of the two
scalar -- neutrino pair diagrams contribute. There are six diagrams
with symmetry breaking vertices: FSSS, FAAS,
FAAA, FASS, FFSS and FFAA; these are again neglected because their are not
HTL's; three non-HTL diagrams of the symmetric theory: FAAS, FSS and FAA; and
four diagrams of the symmetric theory containing small Yukawa couplings:
FASS, FFAS, FFFS and FFFA.

Finally, none of the three possible diagrams FFSS, FFAS and FFAA contributes
to the effective four neutrino vertex \cite{BP2}.
Thus we are left with the effective
contributions to the self-energy of the left-handed neutrino of Fig. 1 a.

Similar considerations of the effective vertices entering to the calculation
of the self-energy of the right-handed neutrino are much more simpler
because the particle is a gauge singlet: there are no three-point or four-point
functions of right-handed neutrino pair and gauge bosons.

The three-point function of a right-handed neutrino pair and a scalar
has two possible one loop contributions:  the diagrams FFS and FSS. The
latter is not HTL by the argument applied already in the
left-handed case. The former is neither HTL: the fermion propagators
produce a term $K^2$ to the denominator of the integrand of the diagram
and hence cancel corresponding factor in the nominator. Effectively
this gives an extra factor of $g$ to the order this diagram.
Thus the effective three point function of the right-handed neutrino pair
and a scalar consists just of the tree level term.

The four-point function of a right-handed neutrino pair and two scalars
has four possible one loop contributions:  the diagrams FSSS, FSS, FFSS and
FFFS. The first two of these are non-HTL diagrams by power counting, while
the third and the fourth are not HTL diagrams because of  the $K^2$ term
in the denominator due to the fermion propagators.

Finally, similarly as in the case of the case of left-handed neutrinos the
possible contribution FFSS to the four neutrino vertex is
not a HTL. Thus we are left with the effective contribution to the
self-energy of the right-handed neutrino of Fig. 1 b.  The effective
neutrino pair -- scalar
three point function consists only of the tree level contribution.

The two-point functions which contribute to the effective propagators
of left- and right-handed neutrinos, scalars and gauge bosons are selected
correspondingly. It turns out that only the diagrams of the symmetric theory
contribute in leading order. Further, in the case of left-handed fermions,
because of the
negligible Yukawa couplings, only the diagram of the type FA contributes
and, in the case of right-handed neutrinos, being gauge singlets, only the
diagram FS contributes. In the case of scalars the two-point function
is independent of the external momentum and it just gives a contribution
to the mass term of the scalars.

The effective, high-temperature, two-point functions  of the gauge bosons
and fermions are well known (see e.g. Refs. \cite{weldon1, weldon2}).
They are listed in the Appendix.

\Section{damping rate}{Zero momentum damping rate}

The neutrino damping rates at zero momentum are calculated to leading
order by evaluating the one-loop diagrams of Figs. 1 a and 1 b, where the
effective
couplings and propagators include the hard thermal loops discussed in
the previous Section. We will consider the damping rates of the
left-handed neutrinos and right-handed neutrinos, that is, of the light
and heavy Majorana neutrinos, separately. There is an essential difference
between these two cases because the interactions of the two sets of
particles differ.

\subsection{Left-handed neutrinos}

In the left-handed sector, according to the analysis of the previous
Section, only the graphs of unbroken gauge theory contribute. Thus the
calculation of effective
$n$-point functions proceeds similarly  as in the case of QCD \cite{BP2},
except that
the group theoretical factors are replaced by  those  corresponding to the
$SU(2)\times U(1)$ symmetry, and one has to take into account the
left-handed nature of  neutrinos, as well as the scalar doublet
contribution to the  gauge boson thermal masses. Although the
modifications to the QCD results are quite straightforward, we present for
completeness some  technical details  in Appendix.

We wish to study the leading symmetry breaking effects, that is, an
inclusion of the vacuum mass terms of gauge bosons, to the zero momentum
damping rate of light neutrinos.  Although the  diagrams arising from the
spontaneous symmetry breaking  were neglected as subleading when   the
effective vertices were calculated, the gauge boson and fermion masses
created through the symmetry breaking, so-called vacuum masses, may have
as important effects to the damping rate  as the thermal masses arising
from the interactions with the plasma.   One should note that the
inclusion of the vacuum mass terms does not  spoil the gauge invariance of
the damping rates, since the Ward identities obeyed by  the
$n$-point functions remain unaltered \cite{BP1}.
These identities
guarantee that the gauge dependent piece of the gauge boson
propagator gives zero contribution to the on-shell damping rate.
The relevant identities are displayed in Appendix.

The effective self-energy of a left-handed neutrino may be written as
\bea{effse}
\Sigma^{(1)}(P) & =T\sum_{k_0}\int\frac{d^3k}{(2\pi)^3}
\left[\Gamma_\mu^A(P,P-K)\Delta_l(P-K)\Gamma^B_\nu(P,P-K)G_{AB}^{\mu\nu}(K)
\right. \nonumber \\ & \qquad \qquad + \left.\frac
12\Gamma^{AB}_{\mu\nu}(P,K)G_{AB}^{\mu\nu}\right].
\eea
Here $\Gamma_\mu^A$ is the effective gauge boson -- neutrino pair three-point
function,
$\Gamma_{\mu\nu}^{AB}$ is the effective four-point vertex for neutrinos
and gauge bosons, $\Delta_l$ is the effective propagator of a left-handed
neutrino, and
$G_{AB}^{\mu\nu}$ is the effective propagator of gauge bosons. The indices
$A,B$ refer to the $SU(2)\times U(1)$ gauge bosons.

The gauge boson propagator may be divided into longitudinal and transverse
projections \cite{weldon1}:
\be{gprop} G^{AB}_{\mu\nu}(K)=\Delta_T^{AB}\hat P_{\mu\nu}+\Delta_L^{AB}\hat
Q_{\mu\nu},
\ee
where $\hat P_{\mu\nu}$, $\hat Q_{\mu\nu}$ together with
$\hat K_{\mu\nu}=k_\mu k_\nu/K^2$ form a set of projection operators
obeying $\hat P_{\mu\nu}+\hat Q_{\mu\nu}+\hat K_{\mu\nu}=g_{\mu\nu}$.
Explicitly,
\be{proj}
\hat P_{\mu\nu}=g_{\mu\nu}-U_\mu U_\nu+
\frac 1{k^2}(K_\mu-k_0U_\mu)(K_\nu-k_0U_\nu),
\ee
where $U=(1,0,0,0)$ is the four-velocity  of the plasma. The gauge
dependent part, which is proportional to $\hat K_{\mu\nu}$,  is not
displayed in \eq{gprop}. The longitudinal and transverse propagators
$\Delta_{L,T}$ have the form
\be{propLT}
\Delta_{L,T}^{AB}(k) = \left(K^2-M^2-\Pi_{L,T}(K)\right)^{-1}_{AB},
\ee
where $M^2$ is the vacuum mass matrix, and $\Pi_{L,T}$ are the
longitudinal and transverse projections of the vacuum polarization tensor.

The damping rate is defined as the imaginary part of the effective
self-energy, which one finds as the solution of the  dispersion relation.
It reads at zero momentum as follows:
\bea{gamma0}
\gamma(0) & = & \frac 18 \mbox{Tr}\left(\gamma_0\,\mbox{Im}\,\Sigma^{(1)}
(p_0=\omega_l,\vek{p}=0)\right) \nonumber \\
          & = & \frac
1{16i}\,\mbox{Disc}\,\mbox{Tr}\left(\gamma_0\Sigma^{(1)}
(p_0=\omega_l,\vek{p}=0)\right).
\eea
The operation $\mbox{Disc}\, f(x)\equiv f(x+i0^+)-f(x+i0^-)$ extracts
the disconnected part of the function $f$, and
$\omega_l$ is the thermal mass of the left-handed neutrino. After
some tedious algebra, we may write the damping rate to the form
\be{gamma02}
\gamma(0)=\frac 1i\,\mbox{Disc}\, T\sum_{k_0}\int\frac{d^3k}{(2\pi)^3}
\left[ \frac 12\{F_A,F_B\}\Delta_T^{AB}\Gamma_T+
       \frac 12\{F_A,F_B\}\Delta_L^{AB}\Gamma_L\right],
\ee
where $F_A$ and $F_B$ are group generators (multiplied with the corresponding
gauge coupling constant) and
\bea{gammatl}
\Gamma_L & = & -\frac{K^2}{k^2}\frac 1{\omega_l^2}\frac 12
\sum_\pm\Delta_\pm(P-K)\left(\omega_l+p_0-k_0\mp k\right)^2, \nonumber \\
\Gamma_T & = & -\frac 1{4\omega_l^2}\sum_\pm\Delta_\pm(P-K)\frac 1{k^2}
\left[\left(p_0-k_0\right)^2-\left( k\pm \omega_l\right)^2\right]^2
\nonumber \\ & &  +\frac
1{4k}\left(1-\frac{(p_0-k_0)^2}{k^2}\right)\ln\left(
\frac{p_0-k_0-k}{p_0-k_0+k}\right).
\eea
The neutrino propagator is here divided into the helicity components
$\Delta_\pm$,
\be{hc}
\Delta_l(P)=\frac 12\Delta_+(\gamma^0-\hat{\vek{p}}\cdot\mbox{\boldmath
$\gamma$})+\frac 12
\Delta_-(\gamma^0+\hat{\vek{p}}\cdot\mbox{\boldmath $\gamma$}).
\ee
The form \eq{gamma02} is similar to the QCD case \cite{BP1}, only that the
gauge boson propagator factors have now a more complicated form.

The sum over the discrete loop energy values may be performed with the
help of the formula
\bea{discformula}
& & \mbox{Disc}
\,T\sum_{k_0}B(k_0)F(p_0-k_0)_{|_{p_0=E}} =
\nonumber \\ & & 2\pi i\left(e^{E/T}+1\right)
\int_{-\infty}^{\infty}d\omega f_B(\omega) \rho_B(\omega)
\int_{-\infty}^{\infty}d\omega' f_F(\omega')\rho_F(\omega')
\delta(E-\omega-\omega'),
\eea
where the argument of the function $B$ ($F$) is even (odd) multiple of
$\pi T i$,
$f_{B(F)}$ is the Bose-Einstein (Fermi-Dirac) distribution,
$\rho_B(\omega)=1/(2\pi i) \mbox{Disc}\, B(\omega)$ is the spectral
function  of $B(\omega)$ and  $\rho_F$ is similarly the spectral
function of $F(\omega)$.

The damping rate of the left-handed neutrino in the zero momentum limit
is thus given by the formula
\be{ldamp}
\gamma(0)=\frac {g_2^2T}{16\pi}\,a,
\ee
where
\bea{a}
a & = & \frac {e^{\omega_l/T}+1}{\omega_l^2}\int k^2dk\,d\omega
\,d\omega'\delta(\omega_l-\omega-\omega')  f_F(\omega')
{\omega\over T}f_B(\omega)\times
\nonumber \\
 & & \left[\frac{2\rho_L(\omega,k)}{\omega}
\sum_\pm\rho_\pm(\omega',k)\left(\omega_l+\omega'\mp k\right)^2 \right. +
\frac{\rho_T(\omega,k)}{\omega}\times
\nonumber \\ & &\left. \left ( \sum_\pm\rho_\pm(\omega',k)
\frac 1{k^2}\left ( \omega'^2-(k\pm \omega_l)^2\right)^2 +
\frac{k^2-\omega'^2}{k^3}\omega_l^2\theta(k^2-\omega'^2)\right)\right],
\eea
and
\bea{disps}
\rho_L & = & \frac{\mbox{Disc}}{2\pi i}\left[\frac 1{2g_2^2}
\{F_A,F_B\}\Delta_L^{AB}(\omega,k)\frac{\omega^2-k^2}{k^2}\right],
\nonumber \\
\rho_T & = & \frac{\mbox{Disc}}{2\pi i}\left[\frac 1{2g_2^2}
\{F_A,F_B\}\Delta_T^{AB}(\omega,k)\right],
\nonumber \\
\rho_\pm & = & -\frac{\mbox{Disc}}{2\pi i}
\left[\Delta_\pm(\omega',k)\right].
\eea
Here $g_2$ is the $SU(2)$ gauge coupling.
Note that \eq{a} applies for the both components of the left-handed lepton
doublet, the neutrino and the charged lepton, by choosing the appropriate
$T_3$ quantum number.

The gauge boson spectral functions separate naturally to the
W-boson, Z-boson and photon terms:
\be{sfb}
\frac{\rho_{L,T}(\omega,k)}{\omega}=
\sum_{K=\gamma,Z,W} Z_{L, T}^K(k)\left(\delta(\omega-\omega_{L,T}^K(k))+
\delta(\omega-\omega_{L,T}^K(k))\right)
+\theta(k^2-\omega^2)R_{L,T}(\omega,k),
\ee
where the delta-function terms arise from the poles of the
propagators $\Delta_{L,T}$ and the step function term from the cut
$k^2>\omega^2$
of the logarithmic terms in the propagators. The explicit expressions of
these are
straightforward to calculate, resulting, however, in rather lengthy formulas
not displayed here.
Similarly, the fermion spectral functions $\rho_\pm$ may be expressed as
\be{sff}
\rho_\pm(\omega,k)=Z_+(k)\delta(\omega\mp\omega_+(k))
+Z_-\delta(\omega\pm\omega_-(k)+\theta(k^2-\omega^2)R_\pm(\omega,k)
\ee
with $R_-(\omega,k)=R_+(-\omega,k)$.

Following Ref. \cite{weldon2}, the residues $Z_{L,T}^{\gamma,Z,W}$ and
$Z_\pm$ may be interpreted as probabilities of creating an excitation
obeying the energy-momentum dispersion relation
$\omega=\omega_{L,T}^{\gamma,Z,W}(k)$ and $\omega=\pm\omega_\pm(k)$ in
the plasma, respectively. In the case of fermions we may interpret the two
branches $\pm\omega_\pm$ as particle-like and hole-like excitations. The
negative-energy solutions are interpreted as antiparticles as
usual. Further, the coefficient of the step function may be
interpreted as the contribution of the particles in the heat bath to
the spectral functions: the quasiparticles are not strictly on mass
shell, because they interact with the heat bath.

With these interpretations, one may divide the coefficient $a$ into
separate contributions according to whether the energies $\omega$ and
$\omega'$ in \eq{a} are off-shell or on-shell. Particularly, if the
both energies are on-shell, the corresponding contributions are interpreted as
processes $\nu_p\, l_h\rightarrow A$ and $\nu_p\, \bar{l}_p\rightarrow
A$, where $A$ denotes generically one of the bosons $\gamma$, $Z$ or
$W$, $l$ denotes generically the left-handed doublet lepton, and
subscripts $p$ and $h$ refer to particle and hole excitations,
respectively \cite{weldon3}. The other possible processes are
kinematically forbidden,
and it is easy to show that the processes above are allowed only if
the thermal mass of the boson is larger than $2\omega_l$. Note that
the hole excitation has same quantum numbers as the anti-particle
excitation but the opposite helicity \cite{weldon2}.

In addition to the on-shell contribution there are contributions where
at least one of the particles in the loop in Fig. 1 a is
off-shell. These contributions may be interpreted as due to the
interactions of the quasiparticles with the heat bath.
The behaviour of the integrand in \eq{a} for these contributions is such that
the main contribution to $a$ comes from the soft region. In that case
one can approximate
$\exp(\omega_l/T)\simeq 1$,
$f_B(\omega)\simeq T/\omega$ and $f_F(\omega')=1/2$. The result is that
the parameter $a$ is independent of temperature (the same is the case in QCD
\cite{BP2}).
The on-shell contribution is exceptional: schematically it reads
\be{aos}
a_{\rm OS}=\frac{Z_{L,T}(k_0)Z_\pm(k_0)}{|\omega_\pm'(k_0)-\omega_{L,T}'|}
f_F(-\omega_\pm(k_0))f_B(\omega_L,T(k_0))\times (\dots),
\ee
where $k_0$ is the solution to the equation $\omega_l+\omega_\pm(k)-
\omega_{L,T}(k)=0$, prime denotes derivation and the ellipsis
represent contributions that are
not relevant in our reasoning. It can be shown that $Z_T$ and $Z_+$ approach
unity when $k$ increases, while $Z_L$ and $Z_-$ approach zero exponentially.
Further, $\omega_T$ and $\omega_+$ approach the free particle dispersion
relation $\omega=(k^2+M^2)^{1/2}$, $M$ is the vacuum mass, with increasing
$k$, while $\omega_L$ and $\omega_-$ approach the massless dispersion relation
$\omega=k$ exponentially. Consider the case of transverse boson and
particle-like neutrino excitations in \eq{aos}. If the vacuum mass of the
gauge boson is large, $M\gg \omega_l$, then $k_0\approx M^2/(2\omega_l)$ and
\be{aos2}
a_{\rm OS}\sim \frac{M^2}{\omega_l^2}e^{-k_0/T},
\ee
which means that without Boltzmann suppression $a$ would grow to high values
with increasing $f\sim M$.

\subsection{Right-handed neutrinos}

The calculation of the zero momentum damping rates of right-handed
neutrinos proceeds along the same lines as of the left-handed ones, the
main difference being the inclusion of the vacuum mass term to the
propagator of the right-handed neutrino.
According to the analysis of the previous section  we need to calculate
the disconnected part of the diagram  in Fig. 1 b. In the case of right-handed
neutrinos the three-point vertex is bare; there is no HTL-contribution to
it. Furthermore, as mentioned in the previous Section, the HTL-contributions
to the scalar
self-energies are independent of the external soft momentum $P$. They
merely add a contribution to the mass term of scalars, which is, in fact,
already taken into account when we use the temperature corrected scalar
potential.

The evaluation of the effective self-energy $\Sigma_R$ of the right-handed
neutrinos follows the same lines as that of the left-handed neutrinos. The
only change is the replacement of the squared thermal masses with
\be{rmat}
\omega^2_r = \lam_M\lam_M^\dagger \frac{T^2}{16}.
\ee
Because we assume that the Dirac mass terms connecting the left- and
right-handed neutrinos are neglected as compared with the Majorana mass
terms, the vacuum mass terms and the thermal energies are diagonalized
simultaneously. The effective lagrangian can then be presented in the
mass  eigenstate basis as follows:
\be{majlag}
{\cal L}_{eff} = \frac 12 \conj{\chi}_l\left(\gamma\cdot P -
\Sigma_L\right)\chi_l+\frac 12 \conj{\chi}_h
\left(\gamma\cdot P -\Sigma^d_R-m_M\right)\chi_h + \cdots
\ee
Here $\chi_l$ and $\chi_h$ represent the light (mainly left-handed)
and heavy (mainly right-handed) mass eigenstates of neutrinos,
respectively. The self-energies $\Sigma_L$ and
$\Sigma_R^d$ are diagonal. Thermal masses of the the heavy flavour
eigenstate neutrinos are (i=1,2,3)
\be{mrd}
m_{R,i}^2(T)=\frac{T^2}{32\bar f^2}m_{M,i}^2.
\ee
Here $m_{M,i}$ is the vacuum mass given by $\sqrt{2}\bar
f\lambda_{M,ii}^{\rm diag}$, where $\lambda_{M}^{\rm diag}$ is the
diagonalized form of the coupling matrix of the right-handed neutrinos and
the singlet scalar defined in Eq. \rf{ly}.

The vacuum mass terms of the  scalars give rise to a mixing between the
singlet Higgs and the neutral member of the Higgs doublet. Hence the
scalar propagator appearing in the neutrino self-energy amplitudes is a
$2\times 2$ matrix. The component of this matrix corresponding to the
dominant channel to which heavy neutrinos couple with a large Yukawa
coupling is given by
\be{scpr11}
\frac{\Gamma_{hh}}{\Gamma_{hh}\Gamma_{ss}-\Gamma_{hs}^2}.
\ee
Here $\Gamma_{hh}=K^2-m^2_{hh}(T)$ and $\Gamma_{ss}=K^2-m^2_{ss}(T)$
are the inverse propagators of scalars $h$ and $s$, and $\Gamma_{hs}$ is the
mass mixing term arising from the scalar potential through the symmetry
breaking. Subscripts $h$ and $s$ refer to the neutral,
P-even  components of the  doublet and singlet fields, respectively.
As a matter of fact, one can neglect in (\ref{scpr11}) the scalar mass
mixing term $\Gamma_{hs}$.
As will be seen in the next Section, inclusion of it would play
subdominant role as compared with the much more substantial effects of
the inclusion of the vacuum mass term of the right-handed neutrino.
In addition to the scalar contribution in \eq{scpr11}, there is also a
Majoron contribution.We shall
neglect the thermal mass differences of the Majoron and the other scalars,
and combine the effects of these particles by using a complex scalar field
$S$ to describe the scalar degrees of freedom.

Unlike for the left-handed neutrinos, the zero momentum damping
rates for the particle- and hole-like excitations differ in the case of the
right-handed neutrinos as a result of the
inclusion of the vacuum mass term. Namely, the imaginary part of the
self-energy $\Sigma^{(1)}_R$ at zero spatial momentum has two contributions
$\alpha_1$ and $\alpha_2$ defined through
\be{imr0}
{\rm Im}\,\Sigma_R^{(1)}(p_0=\omega_r^\pm,\vek p=0)=\alpha_1^\pm\gamma^0
+\alpha_2^\pm,
\ee
where
\be{rthmass}
\omega_r^\pm=(m_R(T)^2+\frac 14 m_M^2)^{1/2}\pm \frac 12 m_M,
\ee
are the thermal masses of the particle (+ sign) and the hole (- sign)
excitations in the case of nonvanishing vacuum mass.
 From the zero momentum dispersion relation
\be{rdispr}
\left(p_0-\frac{m_R(T)^2}{p_0}-i\alpha_1\right)^2-
\left(m_M+i\alpha_2\right)^2=0
\ee
one may then deduce that the zero momentum damping rates $\gamma_\pm(0)$
of the right-handed neutrinos read
\be{dpm}
\gamma_\pm(0)=\frac 12(\alpha_1^\pm\pm\alpha_2^\pm).
\ee

Similarly as in the left-handed calculation  the damping rates
can be written in the form
\be{rdamp}
\gamma_\pm(0)= \frac{|\lam_M^{diag}|^2 T}{16\pi} a_\pm,
\ee
where
\bea{ra}
a_\pm &=& 4(e^{\omega_r^\pm}+1)\times \int k^2dk\int d\omega
d\omega'\delta(\omega_r^\pm-\omega-\omega')\nonumber \\
& &
\frac{\omega}T f_B(\omega) f_F(\omega')\frac{\rho_S(\omega,k)}{\omega}
\left(\rho_{r1}(\omega',k)\pm \rho_{r2}(\omega',k)\right).
\eea
Here the scalar spectral function $\rho_S$ has a very simple form:
\be{rhos}
\frac{\rho_S(\omega,k)}{\omega}=\frac 1{2\omega_S^2(k)}
\left(\delta(\omega-\omega_S(k))+\delta(\omega+\omega_S(k))\right),
\ee
where $\omega_S(k)=(k^2+m_S(T)^2)^{1/2}$. The neutrino spectral
function contains two parts, $\rho_{r1}$ and $\rho_{r2}$, corresponding
to the imaginary parts $\alpha_{1,2}$ in \eq{imr0}.
The right-handed neutrino propagator may be written in the form
\bea{rprop}
\Delta_R(P) & = & \left(f(p_0,p)\gamma_0+h(p_0,p)\hat{\vek{p}}
\cdot\vekk{$\gamma$}-m_M\right)^{-1} \nonumber \\
& = & \frac 1{f^2(p_0,p)-h^2(p_0,p)-m_M^2}
\left(f(p_0,p)\gamma_0+h(p_0,p)\hat{\vek{p}}\cdot\vekk{$\gamma$}+m_M\right),
\eea
where functions $f$ and $h$ are defined through Eqs. \rf{fp} and \rf{se}.
One can then write the spectral functions $\rho_{r1}$ and $\rho_{r2}$ as
follows:
\bea{rdisc}
\rho_{r1}(\omega,k) & = & \frac{\mbox{Disc}}{2\pi i}
\left[\frac{f(\omega,k)}{f^2(\omega,k)-h^2(\omega,k)-m_M^2}\right],
\nonumber \\
\rho_{r2}(\omega,k) & = & \frac{\mbox{Disc}}{2\pi i}
\left[\frac{m_M}{f^2(\omega,k)-h^2(\omega,k)-m_M^2}\right].
\eea

Factorization to the helicity components is not possible in the case
of the right-handed neutrino propagator because of the vacuum mass term.
However, as before the propagator has four poles
$\omega=\pm \omega_r^\pm(k)$,
which may be interpreted as dispersion relations of the particle-,
antiparticle-, hole- and antihole-like excitations, but in contrast with
the vanishing vacuum mass limit there
are now two residues corresponding to the two contributions to the
spectral functions $\rho_{r1}$ and $\rho_{r2}$.
Similarly there are also two off-shell contributions proportional to the step
function $\theta(k^2-\omega^2)$.

The on-shell channels contributing the
damping rates are listed in Table 1. Similarly as in the
calculation concerning the left-handed neutrinos, the on-shell contribution
to the parameter $a_\pm$ may be written schematically as
\be{raos}
a_{\pm,\rm OS}=\frac{Z_{1,2}^{\pm}(k_0)}{|\omega_\pm'(k_0)\pm\omega_S'(k_0)|}
f_F(\pm\omega_r^\pm(k_0))f_B(\pm\omega_S(k_0))\times (\dots),
\ee
where $k_0$ is the solution of the equation
$\omega_r^\pm\pm\omega_r^\pm(k)\pm\omega_S(k)=0$. The sign choices
depend on the channel in question. Again, the residue terms $Z_{1,2}$ for
particle excitations only are non-negligible for large $k_0$.
Further, the denominator in \eq{raos} approaches
zero for large $k_0$, if there is a relative minus sign between the two
derivatives $\omega_\pm'(k_0)$ and $\omega_S'(k_0)$.
These two conditions are fulfilled in the case of the particle damping rate
for the channel $\nu_p\bar{\nu}_p\goto S$
and in the case of the hole damping rate for the channels
$\bar{\nu}_hS\goto\nu_p$ and $\bar{\nu}_h\bar{\nu}_p\goto S$.
These channels give the largest contributions to the damping rates,
as is evident from Figures 3 and 4.

\Section{Numerical results}{Numerical results}

The zero temperature damping rates left- and right-handed neutrinos
derived in the previous section  have quite complicated analytic form. In
this section we analyze their properties numerically. Instead of the
damping rates themselves we will present in our plots the coefficients $a$
and $a_{\pm}$ defined in Eqs. (\ref{ldamp}) and (\ref{rdamp}).

In Fig. 2 we have plotted the coefficient $a$, which is proportional to the
damping rate of the left-handed neutrinos, as a function of the vacuum
expectation value $f$ of the doublet Higgs. As can be seen from the
figure, $a$ is essentially independent on
$f$. This result can be
understood due to the fact that, although the vacuum mass does suppress the
damping rate, there is now a new thermalization
channel open, which has no counterpart e.g. in the pure QCD case
\cite{BP2}. The annihilation of the left-handed neutrino with a neutrino
from the heat bath namely gives  a large contribution to the rate. In fact,
its contribution increases with increasing $f$, in contrast with all the
other contributions which decrease. For $f/T \sim 1.4$ it corresponds to
about a half of the whole rate. For very larger values of $f$, however,
also this annihilation contribution starts to diminish due to Boltzmann
suppression.

The corresponding results  for the right-handed neutrinos are presented in
Figs. 3 and 4. The damping rate and thus the parameter coefficients
$a_\pm$ depend in this case  not only on the  Yukawa coupling but also on the
symmetry breaking mass $m_{S,{vac}} \sim
\lambda_M f'$ as well as the on
thermal mass $m_{S,{th}}\sim T$ of the singlet Higgs $S$,
all of which can be viewed as free parameters.  We have presented $a_+$ and
$a_-$ as functions of the scalar mass $ m_S^2 =
m_{S,{vac}}^2 + m_{S,{th}}^2$ for some representative values of the Yukawa
coupling
$g
\lambda_M$ and ratio $ r = m_M/m_R(T)$, where $m_R$ is the thermal mass and
$m_M$ the vacuum mass of  the right-handed neutrino. It is noteworthy that
the ratio $r$ does not depend on  the Yukawa coupling  constant but only
on the ratio $\bar f/T$. It varies in the range from $r=0$ to
$r\simeq 6$, the upper limit corresponding to its value at $\bar f \simeq T$.

The damping rates of holes and particles behave quite differently when the
parameters are varied, the case of holes showing more structure.  The
general trend for holes is that the larger $r$, the  larger is  the damping
rate. This can be seen in Fig. 3 a. However, for each value of $r$ there exists
an interval in the values of the ratio $m_S(T)/m_R(T)$ where the damping
rate is suppressed. E.g., for
$r = 2$ the damping rate is only $\sim 0.1$ when $m_S(T)/m_R(T)$ lies
between values 2 and 3, while outside this range it is some two orders
of magnitude larger. This is due to the  fact that only soft processes are
allowed in this specific parameter region.

Also we find that the damping rate of the holes is quite sensitive to the
value of the Yukawa coupling $\lambda_M$ (Fig. 3 b). The particle damping rate
instead depends essentially only on the ratio $r$. When
$r$ increases the damping rate decreases until a threshold mass is reached,
as  depicted  in Figs. 4 a and 4 b. At that threshold the annihilation channel
opens which drastically increases the interaction rate. However, the value of
$\lambda_M$   does not affect the damping rate but only marginally. Also
for large $r$ the
damping rates for the particle-like excitation are generally smaller than
for the holes. The main reason is that incoming particles tend to scatter
off the holes, and the number  density of the holes in the heat bath is
suppressed compared with the number density of  particles.

\Section{Conclusions}{Conclusions}

In the present paper we have studied the thermalization of low momentum
neutrinos in the heat bath in the framework of the singlet Majoron model.
This is a relevant issue in connection to the  charge transport mechanism
for generating the baryon asymmetry  during the electroweak phase
transition \cite{ckn} when the  reflection of neutrinos from the phase
wall and from the matter in the broken phase is considered.

We found that the damping rate of the left-handed neutrinos  depends  only
on the  temperature corrected vev $f$ of the doublet Higgs field, not on other
parameters of the model. This dependence is not particularly strong, the
coefficient $a$ defined in  Eq. (\ref{ldamp}) varying between $a = 6$
and $a = 3.5$ when $f$ increases from 0 to
$2T$. The result $a\simeq 6$, corresponding to the unbroken gauge symmetry,
is quite close to the value of the corresponding parameter obtained in QCD
for quark damping, $a\simeq 5.7$ (for three flavours)
\cite{BP1}. We found the damping rate of the left-handed neutrinos being
determined solely by diagrams involving gauge bosons. Therefore our result
can, up to some corrections ${\cal O}(1)$, be directly generalized to the
left-handed charged leptons, too. Furthermore, while the result was derived
in the Majoron model, it is  valid for the left-handed leptons of the
Standard Model as well, since the new degrees of freedom the Majoron model
introduces play no role in the leading order calculation.

The damping rates the  right-handed neutrinos, which do not couple to
the gauge fields, are determined by the Yukawa couplings and properties of
the singlet scalar Higgs. Hence the damping rate is more model dependent
than in the case of left-handed neutrinos. Depending on the values of
various parameters, such as the vacuum mass and the vacuum expectation
value of the singlet Higgs, one can find large as well as small damping
rates for the right-handed  neutrinos.

The present analysis has been performed at the electroweak phase transition
temperature $T \simeq 100$ GeV, whereas the QCD calculations of Ref.
\cite{BP1} was performed at QCD phase transition temperature $T = 200$ MeV
\footnote{It is noteworthy that if QCD calculations of Ref. \cite{BP1}
had been performed at
the electroweak phase transition temperature, the effective number of
degrees  of freedom in the heat bath would be increased.}.
On the other hand, in the region where the gauge symmetry is broken, the
mass of the weak gauge bosons
tends to decrease the damping rate. The net effect is that the rough
estimate for the thermalization time, $1/\gamma \sim 0.1({\mbox{GeV}})^{-1}$,
used e.g.  in \cite{maa}, is about correct for the left-handed leptons,
while the thermalization time of the right-handed  neutrinos may differ
from this crucially.

For each particle species the damping rate should  be smaller than the
particle mass. In the opposite case it would be meaningless to speak about
particles traversing  the medium. In the view were the particle mass is
thought to be built up gradually as a result of sequential forward
scatterings, particles would in this case decay faster that they are formed.
For the right-handed neutrinos this means that the mass given by
(\ref{mrd}) has to be smaller than the damping rate (\ref{rdamp}), implying
$\lambda_M a_\pm < 4 \pi$. This cuts out some large values of $a_\pm$ as
unphysical.

A general conclusion of the present paper is that the damping is fairly
strong, not only in QCD,  but also in the electroweak sector.
Hence the reflection probability of the neutrinos may in general be
substantially reduced  which may crucially affect the charge transport
mechanism of the baryon asymmetry generation of the early universe.
However, there is a region in the parameter space where the damping rate is
suppressed allowing the right-handed
neutrinos to penetrate deep into the broken phase.

\newpage
\noindent{\Large{\bf Acknowledgment}}
\vspace{.4 truecm}

This work has been supported by the Academy of Finland and Turun
Yliopistos\" a\" ati\" o.


\renewcommand{\theequation}{A\arabic{equation}}
\setcounter{equation}{0}
\section*{Appendix}
In this Appendix we present the derivation of the
effective vertices needed for the evaluation of the damping rate
of the left-handed neutrinos.

The hard thermal loops which enter into the calculation of the
damping rate of the left-handed neutrinos arise from the sums of the form
\cite{BP1,FT}
\be{htl} H_n^{\mu_1\dots\mu_{n-1}}=\mbox{Tr}\, K^{\mu_1}\cdots
K^{\mu_{n-1}}
\prod_{i=1}^n\left(\left(K+P_i \right)^2-m_i^2\right)^{-1},
\ee where
\be{tr}
\mbox{Tr}=T\sum_{k_0}\int \frac{d^3k}{(2\pi)^3}.
\ee It is understood that the zero component of the momentum
attached to a boson (fermion) line is an even (odd) multiple of $\pi Ti$,
and the momenta $P_i$ are linear combinations of external momenta.

Consider then the effective three-point function of an $SU(2)$ gauge boson
and a  fermion pair. There are three loops contributing: one with two
$SU(2)$  gauge boson propagators and one fermion propagator; one with two
fermion  propagators and one $SU(2)$ gauge boson propagator; and one with
two  fermion propagators and one $U(1)$ gauge boson propagator. Extracting
the HTL contribution and taking into account the group theoretic factors
we obtain in the case of SU(2) gauge bosons
\be{3v2}
\Gamma_2^{\mu,A} = g_2T^A\left(g^{\mu\nu}+4C_F^gH^{\mu\nu}_{3;2}\right)
\gamma_\nu,
\ee
where $T^A$ is an SU(2) generator and $C_F^g=3/4 g_2^2+Y^2/4 g_1^2$ is the
Casimir operator of the
$({\bf 2},Y)$ representation of $SU(2)\times U(1)$ scaled with the appropriate
couplings. In $H^{\mu\nu}$ the first subscript indicates the number of
internal lines and the second one denotes the number of fermion propagators.
The property $H_{n;m}^{\mu\nu }=-H_{n;n-m}^{\mu\nu }$ was used in derivation
of the \eq{3v2}. A similar
formula is obtained for the effective three-point function $\Gamma_1$ of a
fermion pair and a $U(1)$ gauge boson:
\be{3v1}
\Gamma_1^{\mu} = g_1\frac Y2\left(g^{\mu\nu}+4C_F^gH^{\mu\nu}_{3;2}\right)
\gamma_\nu.
\ee

Note that the three-point functions above are of the general form
\be{3v}
\Gamma_A^{\mu} = F_A\left(g^{\mu\nu}+4C_F^gH^{\mu\nu}_{3;2}\right)
\gamma_\nu,
\ee where $F_A$ denotes the generator of the group in question multiplied by
the  corresponding coupling.

Consider then four-point function of two gauge bosons and a fermion pair.
There are three diagrams (+ crossed channels), contributing: those with
one, two and three fermion propagators. Extracting the HTL-contribution,
calculating the group  theoretic factors and treating
the $SU(2)$ and $U(1)$ gauge bosons equally the four-point
function may be cast to the form
\bea{4v}
\Gamma_{AB}^{\mu\nu}&=&-8\left(-[F_A,F_C][F_B,F_C]+F_C\{F_A,F_B\}F_C\right)
H_{4;1}^{\mu\nu\lam}\gamma_\lam\nonumber\\
&\qquad + &8\left([F_B,F_C]F_AF_C + [F_A,F_C]F_BF_C\right)H_{4;2}^{\mu\nu\lam}
\gamma_\lam.
\eea

As will be shown below, when the gauge boson momenta are equal,  which is
the case in \eq{effse}, the HTL-sums are related as
\be{H-12} H_{4;1}^{\mu\nu\lam}=-\frac 12 H_{4;2}^{\mu\nu\lam}.
\ee
Inspection shows that the four-point function can be written in this
case as
\be{4veff}
\Gamma_{AB}^{\mu\nu}=4C_F^g\{F_A,F_B\}H_{4;2}^{\mu\nu\lam}\gamma_\lam.
\ee

The sum in \eq{htl} is most easily evaluated by applying the formula  (see
e.g. \cite{kapusta})
\be{}
T\sum_{k_0}F(k_0)=\frac{\eta}{2\pi i}\oint dz f_\eta(z)[F(z)+F(-z)],
\ee
where $\eta=-1$ corresponds to the Bose-Einstein and $\eta=+1$ to the
Fermi-Dirac  distribution, the appropriate distribution chosen according to
whether $k_0$ is even or odd multiple of $\pi Ti$. The integration path
encloses all the  poles of the function $F(z)$ on the right half of the
$z$-plane. A calculation results in
\bea{htl2} H_n^{\mu_1\dots\mu_{n-1}}&=&
\int \frac{d^3k}{(2\pi)^3}\sum_{j=1}^n  \eta_j
\left\{{\cal K}_{|k_0=\omega_j-p^0_j}\frac{f_{\eta_j}(\omega_j)}{2\omega_j}
\prod_{i\neq j} \left[(\omega_j-p^0_j+p_i^0)^2-\omega_i^2\right]^{-1}
\right.
\nonumber \\ & \qquad + &
\left.{\cal K}_{|k_0=-\omega_j-p^0_j}\frac{f_{\eta_j}(\omega_j)}{2\omega_j}
\prod_{i\neq j} \left[(\omega_j+p^0_j-p_i^0)^2-\omega_i^2\right]^{-1}
\right\}.
\eea Here a shorthand notation ${\cal K}=K^{\mu_1}\cdots K^{\mu_{n-1}}$
was used and
\be{omega}
\omega_i=\sqrt{(\vek{k}+\vek{p}_i)^2+m_i^2}.
\ee Next we perform a linear approximation in external momenta in the
denominators and neglect the dependence of the external momenta elsewhere:
\bea{htl3} H_n^{\mu_1\dots\mu_{n-1}} & = &
\frac 1{\pi^2 2^{n+1}}\sum_{j=1}^n\eta_j\int_0^{\infty}dk\,kf_{\eta_j}(k)
\times
\nonumber \\ &  & \int\frac{d\Omega}{4\pi}
\left\{\prod_{i\neq j}\frac{\hat{{\cal K}}}{(P_i-P_j)\cdot\hat{K}}
 + \prod_{i\neq j}\frac{\hat{{\cal K'}}}{(P_i-P_j)\cdot\hat{K'}}\right\}.
\eea
Here $\hat{K}=(1,+\hat{\vek{k}})$, $\hat{K}'=(1,-\hat{\vek{k}})$, and the
numerator $\hat{{\cal K}}$ is defined as $\hat{{\cal K}}=\hat{K}^{\mu_1}
\cdots\hat{K}^{\mu_{n-1}}$ (and similarly for $\hat{{\cal K}}'$). Finally,
after the $k$-integral is performed we end up with
\be{htlf} H_n^{\mu_1\dots\mu_{n-1}} =
\frac{T^2}{6\cdot 2^n}\int\frac{d\Omega}{4\pi}\hat{\cal K}
\left\{-\sum_{j,B}\prod_{i\neq j}\frac 1{(P_i-P_j)\cdot\hat{K}}
 +\frac 12\sum_{j,F}\prod_{i\neq j}\frac 1{(P_i-P_j)\cdot\hat{K}}\right\}.
\ee Here the sum over internal lines is divided to the fermion and boson
contributions.

Using the \eq{htlf} and plugging in the appropriate momentum configuration
in the function $H^{\mu\nu}_{3;2}$ gives
\be{3vf}
\Gamma^\mu_A = F_A\left(\gamma^\mu+\omega_l^2(T)\int\frac{d\Omega}{4\pi}
\frac{\hat{K}\cdot\gamma\hat{K}^\mu}{\hat{K}\cdot P_1\,\hat{K}\cdot P_2}
\right),
\ee where $P_1$ ($P_2$)is the incoming (outgoing) momentum of the fermion
and
\be{tm} \omega_l(T)^2=C_F^g\frac{T^2}8
\ee is the thermal mass of the left-handed fermion \cite{joku}.

The relevant momentum configuration of the function $H_{4;1}$ is
$P_1=P_2=0$, $P_3=-Q$ and $P_4=P$, where $Q$ is the momentum of the
incoming  and outgoing gauge boson and $P$ is the momentum of the incoming
and outgoing fermion. The term inside the braces in \eq{htlf} may in this
case be expressed as
\be{h41}
\{\dots\}=\frac 32 \frac 1{(\hat{K}\cdot P)^2\,\hat{K}\cdot (P+Q)}.
\ee
On the other hand, the relevant momentum configuration of the function
$H_{4;2}$ is $P_1=0$, $P_2=Q$, $P_3=P$ and $P_4=P+Q$, and the term inside
the braces in \eq{htlf} becomes
\be{h42}
\{\dots\}=-\frac 32 \frac 1{(\hat{K}\cdot P)^2}
\left(\frac 1{\hat{K}\cdot (P-Q)}+\frac 1{\hat{K}\cdot (P+Q)}\right).
\ee
As the momentum $Q$ is integrated over, and the propagator of the gauge boson
is symmetric in $Q$, one can conclude from (\ref{h41}) and (\ref{h42})
that $H_{4;1}=-1/2 H_{4;2}$. The four-point function has then the form
\be{4vf}
\Gamma_{AB}^{\mu\nu}=\omega_l^2(T)\{F_A,F_B\}\int\frac{d\Omega}{4\pi}
\frac{\hat{K}\cdot\gamma\hat{K}^\mu\hat{K}^\nu} {(\hat{K}\cdot
P)^2\,\hat{K}\cdot (P+Q)}.
\ee

The effective propagators may be calculated along similar lines.
The gauge boson vacuum polarization functions read as
\bea{vp}
\Pi_T^{AB} &= \delta^{AB} \Pi_T&=  \delta^{AB} 3M(T)^2(1-x^2)(1-\frac x2L),
\nonumber \\
\Pi_L^{AB} &= \delta^{AB} \Pi_L &=  \delta^{AB} (\frac 32M(T)^2-\frac 12\Pi_T),
\eea
where $x=\omega/k$ and $L=\log((1+x)/(1-x))$. The thermal masses for
the $SU(2)$ and $U(1)$ gauge bosons are up to the coupling constants equal:
\bea{tem} M^2_{SU(N)} & = & \frac{N_f+4N+2}4
\frac{g_2^2T^2}9=\frac{11}{18}g_2^2T^2
\nonumber \\ M^2_{U(1)}  & = & \frac{\sum Y^2_L+\sum Y^2_R+2}8
\frac{g_1^2T^2}9=
\frac{11}{18}g_1^2T^2.
\eea
Here $N_f$ is the number of left-handed $SU(2)$ doublet fermion
representations present in the thermal bath. We have performed the
calculations with $N_f=12$, i.e. the thermal bath is assumed to contain also
top quarks. However, the results are not expected to be very sensitive
to small variations of $N_f$ or $Y_{L,R}$. The contribution
${2\over 8} {g_1^2 T^2\over 9}$ in (\ref{tem}) come from the scalar doublet
contribution.

The propagator sum of the gauge bosons may be cast to the form
\be{bp}
\frac 1{2g_2^2}\{F_A,F_B\}\Delta_{L,T}^{AB} =
\frac 12 \frac 1{\Gamma_2^{L, T}} + \frac 14 (\Gamma_1^{L,T}\Gamma_2^{L,T}-
\Gamma_{12}^2)^{-1}(\Gamma_1^{L,T} + \Gamma_2^{L,T} \tan^2\theta_w +
2 \Gamma_{12}\tan\theta_w),
\ee
where
\bea{kamma}
\Gamma_{1,2}^{L,T} & = & \omega^2-k^2-\Pi^{1,2}_{L,T}-g_{1,2}^2\left(\frac
f2\right)^2,
\nonumber \\
\Gamma_{12} & = & g_1g_2\left(\frac f2\right)^2,
\eea
and $\tan\theta_w = g_1/g_2$.

The fermion propagators are of the form
\bea{fp}
\Delta_L^{-1} & = & \gamma\cdot P-\Sigma_L, \nonumber \\
\Delta_R^{-1} & = & \gamma\cdot P-\Sigma_R-m_M,
\eea
where
\be{se}
\Sigma_{L,R}(P)=\omega_{l,r}^2(T)\int\frac{d\Omega}{4\pi}
\frac{\hat{K}\cdot\gamma}{\hat{K}\cdot P}=\frac{\omega_{l,r}^2}{2p}
\left[L\gamma_0+(2-xL)\hat{\vek{p}}\cdot\vekk{$\gamma$}\right],
\ee
the notation being the same as in \eq{vp}, $\omega_l$ is the thermal mass
from \eq{tem} and $\omega_r^2$ defined in Eq. (\ref{rmat})
gives the thermal mass of the right-handed neutrino arising from the
scalar-fermion loop.

Using Eqs. \rf{3vf}, \rf{4vf}, \rf{vp} and \rf{se}, it is easy to deduce
the following Ward identities:
\bea{award}
(P_1-P_2)_\mu\Gamma_A^\mu(P_1,P_2) & = & F_A\left(\Delta_l(P_1)^{-1}-
\Delta_l(P_2)^{-1}\right) \nonumber \\
Q_\mu Q_\nu\Gamma^{\mu\nu}_{AB}(P,Q) & = & \{F_A,F_B\}\left(\Delta_l(P+Q)^{-1}
-\Delta_l(P)^{-1}\right).
\eea
The latter of these identities holds only in the case where there is an
integration over gauge boson momentum $Q$.


\newpage
\noindent {\Large{\bf Figure captions}}
\vskip .5truecm
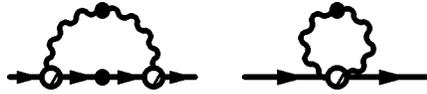
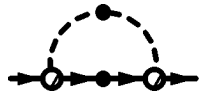
\noindent Figure 1. Generic effective diagrams for the  neutrino damping
rates: a) diagrams contributing to the left-handed neutrinos, b) diagrams
contributing to the right-handed neutrinos and c) non-contributing
Diagrams.
\vspace{0.5cm}

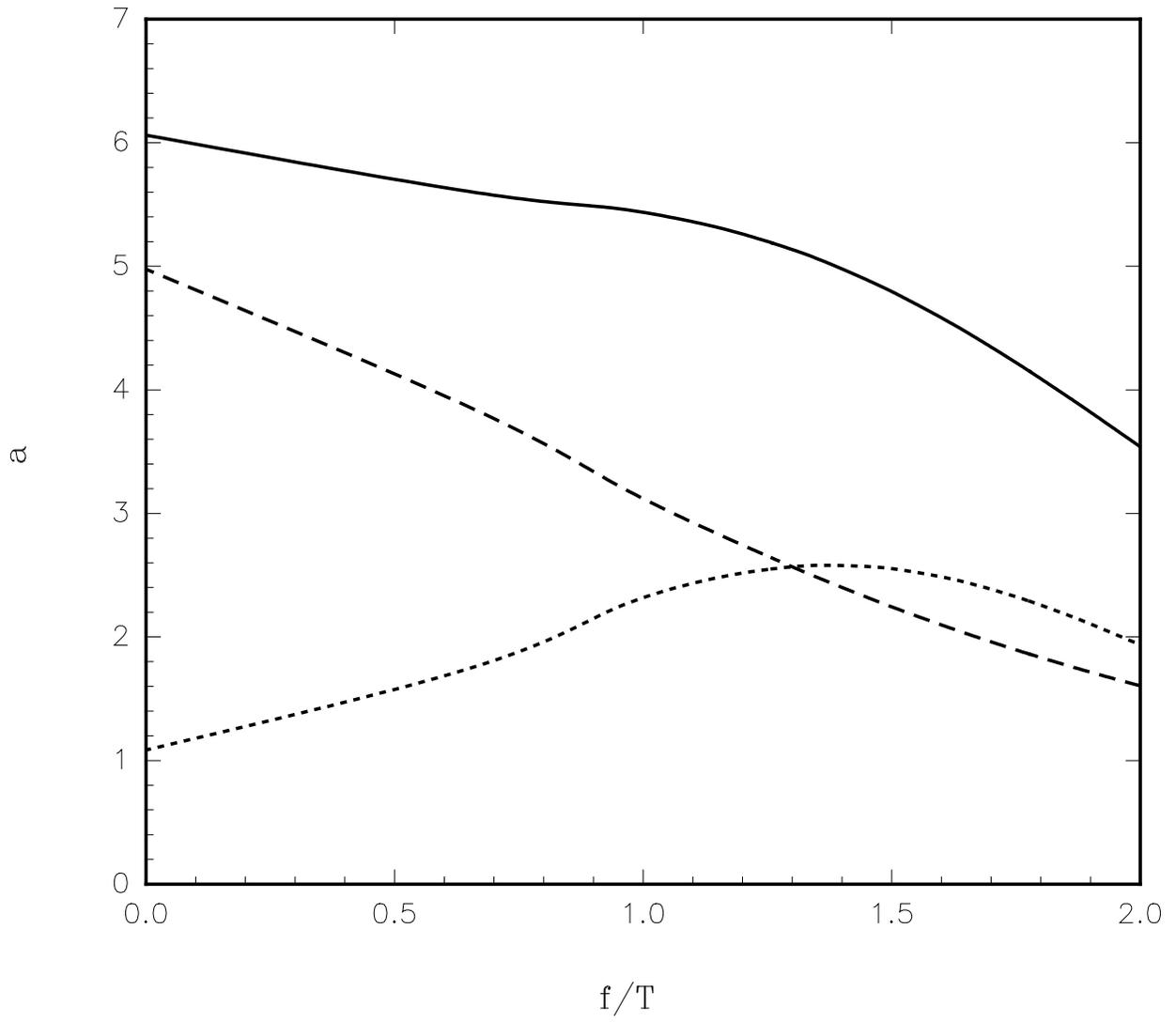
\noindent Figure 2. Damping rate of the left-handed neutrinos as a function
of the vacuum expectation value of the doublet field: the total rate
(solid  line), contribution coming from particle-hole annihilation (dashed
line) and the other contributions (dotted line).
\vskip .5truecm

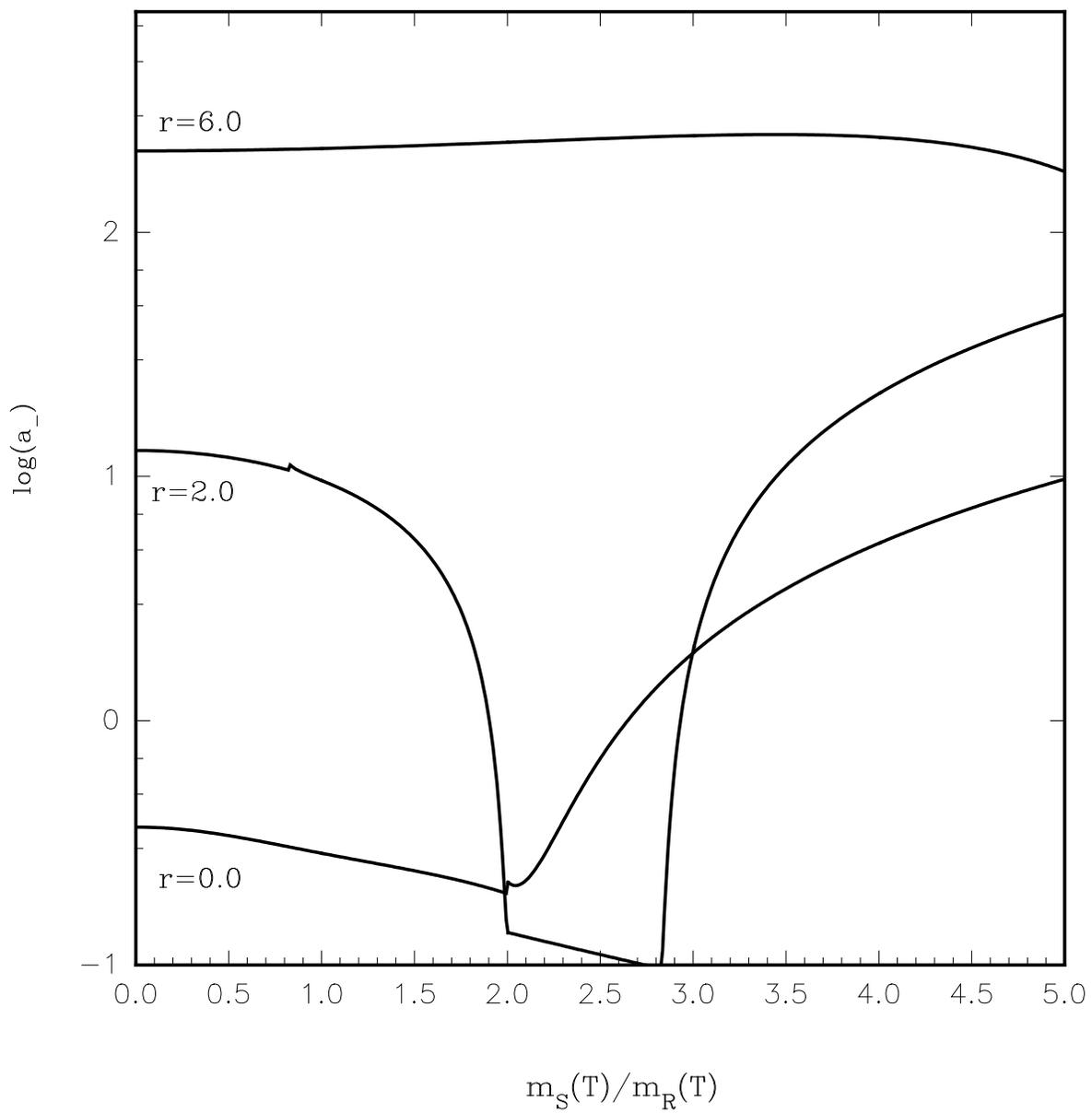
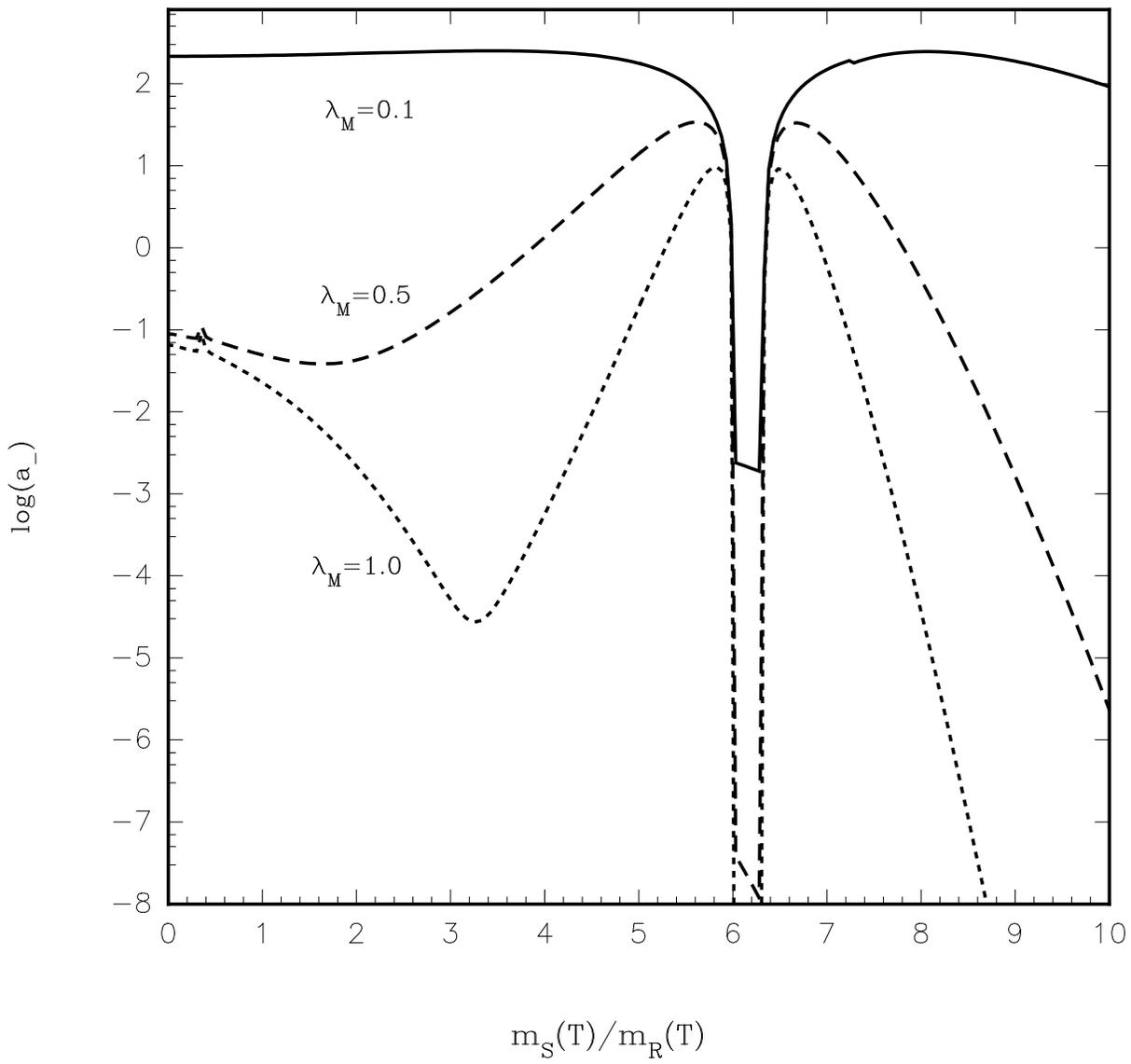
\noindent Figure 3. Damping rates of right-handed hole-like excitations of
neutrinos as function of the total scalar mass $m_S$: (a) For Yukawa
coupling
$\lambda_M = 0.1$ and various ratios $r =$ 0, 2 and 6; (b) For fixed ratio
$r = 6$  and for Yukawa couplings $\lambda_M =$ 0.1, 0.5, 1.
\vskip .5truecm

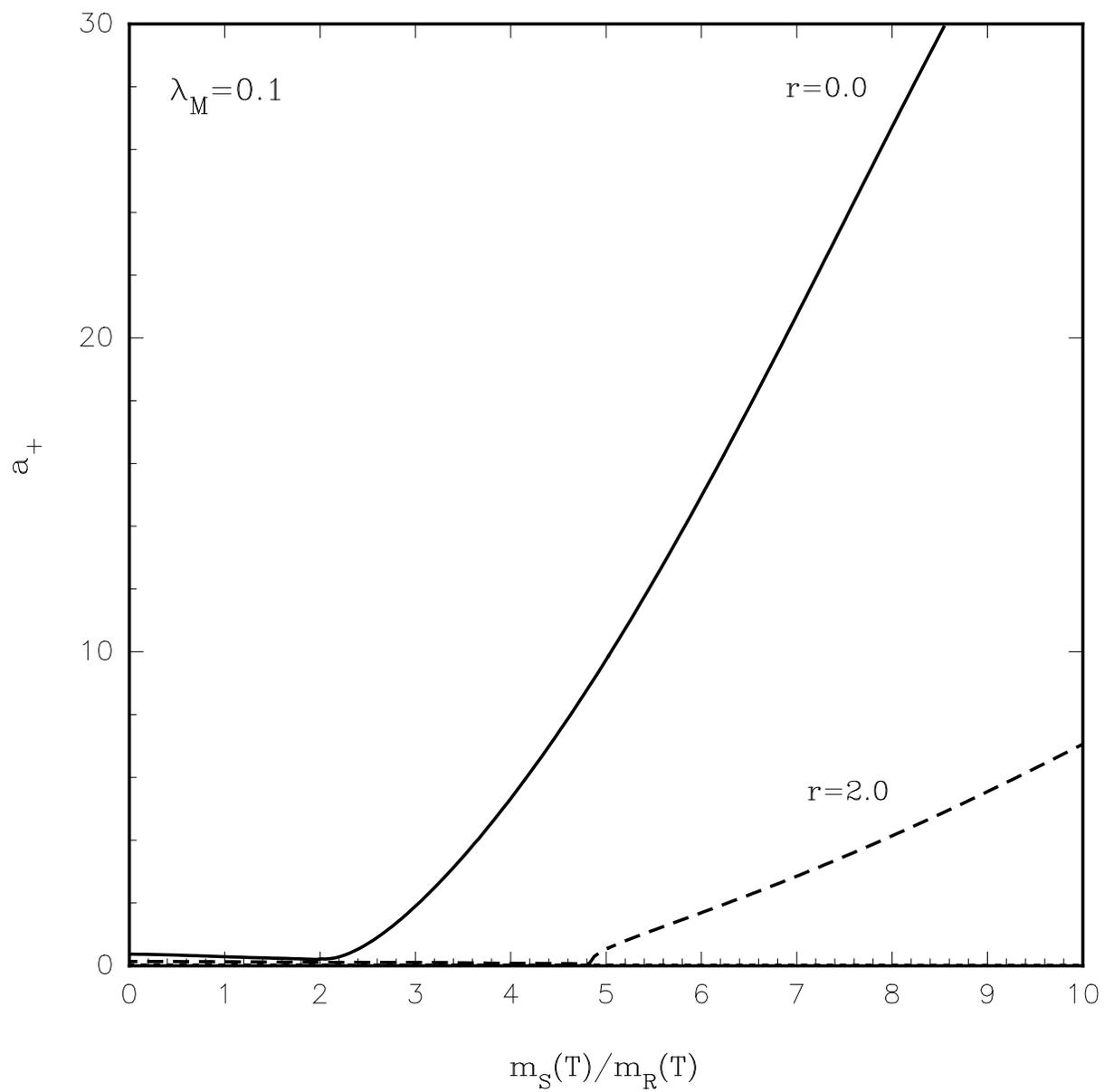
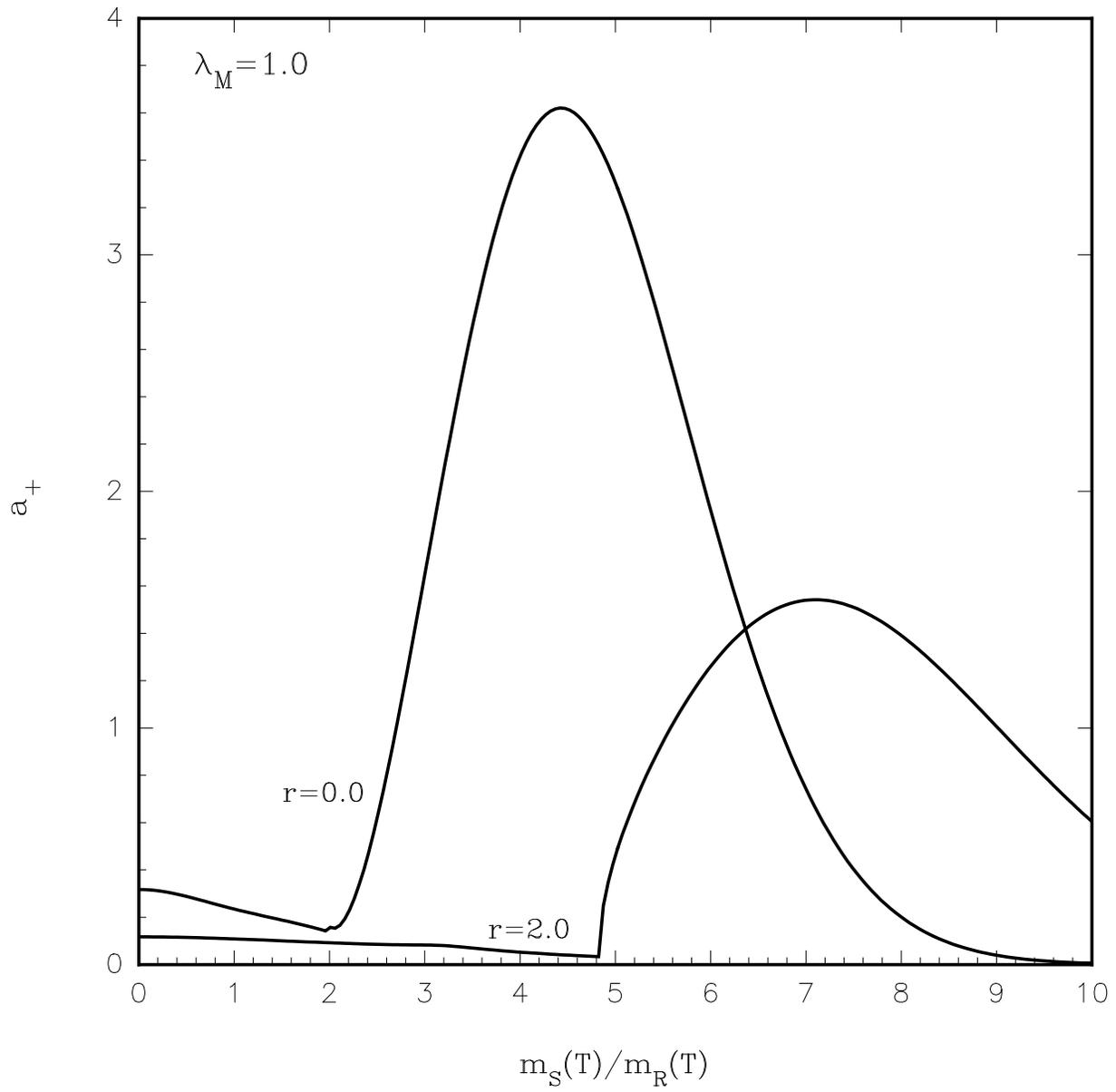
\noindent Figure 4. Damping rates of right-handed particle-like excitations
of neutrinos as function of the total scalar mass $m_S$ with
ratios $r =$ 0, 2: (a) $\lambda_M = 0.1$, (b) $\lambda_M = 1.0$.
\vspace{1cm}

\noindent {\Large\bf Table caption }
\vspace{0.5cm}

\noindent Table 1. The decay channels of the on-shell contribution
to the damping rate of the right-handed particle ($\gamma_+$) and hole
($\gamma_-$) excitations and corresponding mass thresholds.
\newpage
\center{\Huge\bf Table 1}
\vspace{2cm}
\center{\large
\begin{tabular}{lc|c|c}  \cline{2-4}
$\gamma_+$ & $ m_S<\orp-\orm $ & $ m_S>\orp-\orm $ & $ m_S>2\orp $ \\
\cline{2-4}
 & $\nu_p\goto S\bar{\nu}_h$&$\nu_p\nu_h\goto S$&$\nu_p\bar{\nu}_p\goto S$ \\
\end{tabular}}
\vspace{0.5cm}
\center{\large
\begin{tabular}{lc|c|c} \cline{2-4}
$\gamma_-$ & $ m_S<\orp-\orm $ & $ m_S>\orp-\orm $ & $ m_S>2\orm $ \\
\cline{2-4}
 & $\bar{\nu}_hS\goto\nu_p$&$\bar{\nu}_h\bar{\nu}_p\goto S$&$\nu_h\bar{\nu}_h
\goto S$ \\
\end{tabular}}

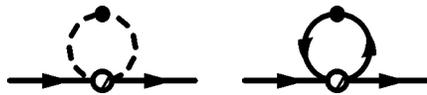
\end{document}